\documentclass[nofootinbib,prd,showpacs]{revtex4}%
\usepackage{amsmath,amsfonts,amssymb,amscd,graphicx}
\catcode`\@=11
\catcode`\@=12

\newfont{\gotico}{eufm10 scaled\magstephalf}
\newfont{\qvd}{msam10 scaled\magstephalf}

\def\de#1/de#2{\frac{\partial {#1}}{\partial {#2}}}
\def\De#1/de#2{\dfrac{\partial {#1}}{\partial {#2}}}

\def\const{{\rm const.}}

\def\beqa{\begin{eqnarray}}
\def\eeqa{\end{eqnarray}}
\def\beq{\begin{equation}}
\def\eeq{\end{equation}}

\renewcommand{\epsilon}{\varepsilon}

\def\bet{\begin{tabular}}
\def\eet{\end{tabular}}
\def\bef{\begin{figure}}
\def\eef{\end{figure}}
\def\beqa{\begin{eqnarray}}
\def\eeqa{\end{eqnarray}}
\def\beq{\begin{equation}}
\def\eeq{\end{equation}}

\def\ie{{\it i.e. }}

\def\p{\phi}

\renewcommand{\epsilon}{\varepsilon}

\def\beqa{\begin{eqnarray}}
\def\eeqa{\end{eqnarray}}
\def\beq{\begin{equation}}
\def\eeq{\end{equation}}

\def\ie{{\it i.e. }}

\def\pr{{\it Phys. Rev.}\ }
\def\prl{{\it Phys. Rev. Lett.}\ }
\def\pl{{\it Phys. Lett.}\ }

\def\cqg{{\it Class. Quantum Grav.}\ }

\def\grg{{\it Gen. Relativ. Grav.}\ }

\def\mnras{{\it Mon. Not. R. Ast. Soc.}\ }

\begin{document}
\vskip-2cm

\title{$f\/(R)$ gravity with torsion: the metric-affine approach}

\author{S. Capozziello$^{1}$, R. Cianci$^{2}$, C. Stornaiolo$^{1}$, S.
Vignolo$^{2}$}

\affiliation{$~^{1}$ Dipartimento di Scienze Fisiche,
Universit\`{a} ``Federico II'' di Napoli and INFN Sez. di Napoli,
Compl. Univ. Monte S. Angelo Ed. N, via Cinthia, I- 80126  Napoli
(Italy)}

\affiliation{$^{2}$DIPTEM Sez. Metodi e Modelli Matematici,
Universit\`a di Genova,  Piazzale Kennedy, Pad. D - 16129 Genova
(Italy)}

\begin{abstract}
The role of torsion in $f(R)$ gravity is considered in the
framework of metric-affine formalism. We discuss the field
equations in empty space and in presence of perfect fluid matter
taking into account the analogy with the Palatini formalism. As a
result, the extra curvature and torsion degrees of freedom can be
dealt as an effective scalar field of fully geometric origin. From
a cosmological point of view, such a geometric description could
account for the whole Dark Side of the Universe.
\end{abstract}


\maketitle

\section{Introduction}
In the last thirty years, some shortcomings came out in the
Einstein General Relativity (GR) and several investigations
started in order to study if alternative approaches to
gravitational interaction are possible and self-consistent. Such
issues  come from Cosmology and Quantum Field Theory. In the first
case, the presence of  Big Bang singularity, flatness and horizon
problems \cite{guth} led to the result that Standard Cosmological
Model \cite{weinberg}, based on GR and Standard Model of particle
physics, is inadequate to describe the Universe at extreme
regimes. On the other hand, GR  does not work for a quantum
description of spacetime. Due to this facts and to the lack of a
Quantum Gravity theory, alternative theories of gravity have been
pursued in order to attempt, at least, a semi-classical scheme
where GR and its positive results could be recovered.

A fruitful approach has been that of {\it Extended Theories of
Gravity} (ETG) which have become a sort of paradigm in the study
of gravitational interaction. They are essentially  based on
corrections and enlargements of the Einstein theory. The paradigm
consists in adding higher-order curvature invariants and
non-minimally coupled scalar fields into dynamics resulting from
the effective action of Quantum Gravity \cite{odintsov,farhoudi}.

Other motivations to modify GR come from the issue to recover
Mach's principle \cite{brans}. This principle states that the
local inertial frame is determined by some average of the motion
of distant astronomical objects \cite{bondi}, so that
gravitational coupling can be scale-dependent and related to some
scalar field. This viewpoint leads to assume a varying
gravitational coupling. As a consequence,  the concepts of
``inertia'' and equivalence principle have to be revised
\cite{brans,cimento,sciama,faraoni}.

Furthermore, every unification scheme as Superstrings,
Supergravity or Grand Unified Theories, takes into account
effective actions where non-minimal couplings to the geometry or
higher-order terms in the curvature invariants come out. Such
contributions are due to one-loop or higher-loop corrections in
the high-curvature regimes. In particular, this scheme has been
adopted in order to deal with  quantization in curved spacetimes
and, as a result, the interactions between quantum scalar fields,
 the gravitational self-interactions and the background geometry yield
correction terms in the Hilbert-Einstein Lagrangian
\cite{birrell}.

Moreover, it has been realized that such  terms are inescapable if
we want to obtain the effective action of Quantum Gravity on
scales closed to the Planck length \cite{vilkovisky}. Higher-order
terms in the curvature invariants, as $R^{2}$, $R^{ij} R_{ij}$,
$R^{ijkh}R_{ijkh}$, $R \,\Box R$, $R \,\Box^{k}R$ or non-minimally
coupled terms between scalar fields and geometry, as $\p^{2}R$,
have to be added to the effective Lagrangian of gravitational
field when quantum corrections are considered. For example, one
has to stress that such terms occur in the effective Lagrangian of
strings or in Kaluza-Klein theories, when the mechanism of
dimensional reduction is used \cite{veneziano}.

From a conceptual point of view, there would be no {\it a priori}
reason to restrict the gravitational Lagrangian to a linear
function of the Ricci scalar $R$, minimally coupled with matter
\cite{francaviglia}. The idea that there are no ``exact'' laws of
physics but that the Lagrangians of physical interactions are
``stochastic'' functions -- with the property that local gauge
invariances (\ie conservation laws) are well approximated in the
low energy limit and that physical constants can vary -- has been
taken into serious consideration -- see Ref.~\cite{ottewill} and
references therein.

Besides fundamental physics motivations, all these theories have
acquired a huge interest in cosmology due to the fact that they
``naturally" exhibit inflationary behaviors able to overcome the
shortcomings of Standard Cosmological Model (based on GR). The
related cosmological models seem very realistic and capable of
matching with the observations \cite{starobinsky,la,kerner}.

Furthermore, it is possible to show that, via conformal
transformations, the higher-order and non-minimally coupled terms
always correspond to  Einstein gravity plus one or more than one
minimally coupled scalar fields
\cite{teyssandier,maeda,wands,gottloeber,magnano}. More precisely,
higher-order terms always appear as a contribution of order two in
the equations of motion. For example, a term like $R^{2}$ gives
fourth order equations \cite{ruzmaikin}, $R \ \Box R$ gives sixth
order equations \cite{gottloeber,sixth}, $R \,\Box^{2}R$ gives
eighth order equations \cite{eight} and so on. By a conformal
transformation, any 2nd-order of derivation corresponds to a
scalar field: for example, fourth-order gravity gives Einstein
plus one scalar field, sixth order gravity gives Einstein plus two
scalar fields and so on \cite{gottloeber,schmidt1}. This feature
results very interesting if we want to obtain multiple
inflationary events since a former early stage could select
``very'' large-scale structures (clusters of galaxies today),
while a latter stage could select ``small'' large-scale structures
(galaxies today) \cite{sixth}. The philosophy is that each
inflationary era is connected with the dynamics of a scalar field.
Furthermore, these extended schemes naturally could solve the
problem of ``graceful exit" bypassing the shortcomings of former
inflationary models \cite{la,aclo}.

Recently, ETG are going also to play an interesting role to
describe the today observed Universe. In fact, the  amount of good
quality data of last decade has made it possible to shed new light
on the effective picture of the Universe.  Type Ia Supernovae
(SNeIa) \cite{SNeIa}, anisotropies in the cosmic microwave
background radiation (CMBR) \cite{CMBR}, and  matter power
spectrum inferred from large galaxy surveys \cite{LSS} represent
the strongest evidences for a radical revision of the Cosmological
Standard Model also at recent epochs. In particular, the {\it
concordance $\Lambda$CDM model} predicts that baryons contribute
only for $\sim 4\%$ of the total matter\,-\,energy budget, while
the exotic {\it cold dark matter} (CDM) represents the bulk of the
matter content ($\sim 25\%$) and the cosmological constant
$\Lambda$ plays  the role of the so called "dark energy" ($\sim
70\%$) \cite{triangle}. Although being the best fit to a wide
range of data \cite{LambdaTest}, the $\Lambda$CDM model is
severely affected by strong theoretical shortcomings
\cite{LambdaRev} that have motivated the search for alternative
models \cite{PR03,copeland}.

Dark energy models mainly rely on the implicit assumption that
Einstein's GR is the correct theory of gravity indeed.
Nevertheless, its validity at the larger astrophysical and
cosmological scales has never been tested \cite{will}, and it is
therefore conceivable that both cosmic speed up and dark matter
represent signals of a breakdown in our understanding of
gravitation law so that one should consider the possibility that
the Hilbert\,-\,Einstein Lagrangian, linear in the Ricci scalar
$R$, should be generalized.

Following this line of thinking, the choice of a generic function
$f(R)$ can be derived by matching the data and by the "economic"
requirement that no exotic ingredients have to be added. This is
the underlying philosophy of what is referred to as $f(R)$ gravity
\cite{capozzcurv,odi1,odi2,curvrev,cdtt,cdtt2,flanagan,allemandi,odintsovfr,mimicking,koivisto}.
In this context, the same cosmological constant could be removed
as an ingredient of the cosmic pie being nothing else but a
particular eigenvalue of a general class of theories
\cite{garattini}.

However $f(R)$ gravity can be encompassed in the ETGs being a
"minimal" extension of GR where (analytical) functions of Ricci
scalar are taken into account.

Although higher order gravity theories have received much
attention in cosmology, since they are naturally able to give rise
to  accelerating expansions (both in the late and in the early
Universe) and systematic studies of the phase space of solutions
are in progress \cite{cnot,tsu2,tsu3,noi-phase,ctd}, it is
possible to demonstrate that $f(R)$ theories can also play a major
role at astrophysical scales. In fact, modifying the gravity
Lagrangian can affect the gravitational potential in the low
energy limit. Provided that the modified potential reduces to the
Newtonian one on the Solar System scale, this implication could
represent an intriguing opportunity rather than a shortcoming for
$f(R)$ theories (see for example
\cite{anderson,farasis,faulkner,bertolami,odisis}).

Furthermore, a corrected gravitational potential could offer the
possibility to fit galaxy rotation curves without the need of dark
matter \cite{noipla,mond,jcap,salucci}. In addition, it is
possible to work out a formal analogy between the corrections to
the Newtonian potential and the usually adopted dark matter
models. In general, any relativistic theory of gravitation can
yield corrections to the Newton potential \cite{schmidt}) which,
in the post-Newtonian (PPN) formalism, could give rise to tests
for the same theory \cite{will,ppnantro,arturo,matteo}.

In this paper, we want to face the problem to study $f(R)$ gravity
considering also torsion. Torsion theories have been taken into
account firstly by Cartan and then where introduced by Sciama and
Kibble in order to deal with spin in General Relativity (see
\cite{hehl} for a review). Being the spin as fundamental as the
mass of the particles, torsion was introduced in order to complete
the following scheme: the mass (energy) as the source of curvature
and the spin as the source of torsion.

Up to some time ago, torsion  did not seem to produce models with
observable effects since  phenomena implying  spin and gravity
were considered to be significant only in the very early Universe.
After, it has been proven  that spin  is not the only source of
torsion. As a matter of fact, torsion field can be decomposed in
three irreducible tensors, with different properties. In
\cite{Capozziello:2001mq}, a systematic classification of these
different types of torsion and their possible sources is
discussed. This means that a wide class of torsion models could be
investigate  independently of spin as their source.

In principle, torsion could be constrained at every astrophysical
scale and, as recently discussed, data coming from Gravity Probe B
could contribute to this goal also  at Solar System level
\cite{tegmark}.

In \cite{Sotiriou-Liberati1,Sotiriou-Liberati2}, a systematic
discussion of metric-affine $f(R)$ gravity has been pursued. In
particular,  the role of connection  in presence of matter has
been studied considering the various possible matter actions
depending on connection. The main result of these papers has been
the evidence that matter can tell to spacetime how to curve as
well as how to twirl.

In this paper, following the same philosophy, we want to show
that, starting from a generic $f(R)$ theory, the curvature and the
torsion can give rise to an effective curvature-torsion
stress-energy tensor capable, in principle, to address the problem
of the Dark Side of the Universe in a very general geometric
scheme. We do not consider the possible microscopic distribution
of spin but a general  torsion vector field in $f(R)$ gravity.

The layout of the paper is the following. In Secs.II and III, we
derive the metric-affine field equations of $f(R)$ gravity with
torsion in empty space and in presence of matter, respectively.
Sec.IV is devoted to the discussion of the formal equivalence with
scalar-tensor theories, while applications to
Friedmann-Robertson-Walker (FRW) cosmology are discussed in Sec.V.
Summary and conclusions are drawn in Sec. VI.

\section{Field equations in empty space}

Let us discuss the main features of a $f(R)$ gravity considering
the most general case in which torsion is present in a ${\cal
U}_4$ manifold. In a  metric--affine formulation, the metric ${
g}$ and the connection ${ \Gamma}$ can be, in general, considered
independent fields. More precisely, the dynamical fields are pairs
$(g,\Gamma)\/$ consisting of a pseudo--Riemannian metric $g\/$ and
a metric compatible linear connection $\Gamma\/$ on the
space--time manifold $M\/$. The corresponding field equations are
derived by varying separately with respect to the metric and the
connection the action functional
\begin{equation}\label{00.1}
{\cal A}\/(g,\Gamma)=\int{\sqrt{|g|}f\/(R)\,ds}
\end{equation}
where $f\/$ is a real function, $R\/(g,\Gamma) = g^{ij}R_{ij}\/$
(with $R_{ij}:= R^h_{\;\;ihj}\/$) is the scalar curvature
associated with the connection $\Gamma\/$ and $ds :=
dx^1\wedge\dots\wedge dx^4\/$. Throughout the paper we use the
index notation
\begin{equation}\label{00.1bis}
R^h_{\;\;kij}=\de{\Gamma_{jk}^{\;\;\;h}}/de{x^i} -
\de{\Gamma_{ik}^{\;\;\;h}}/de{x^j} +
\Gamma_{ip}^{\;\;\;h}\Gamma_{jk}^{\;\;\;p} -
\Gamma_{jp}^{\;\;\;h}\Gamma_{ik}^{\;\;\;p}
\end{equation}
for the curvature tensor and
\begin{equation}\label{00.1tris}
\nabla_{\de /de{x^i}}\de /de{x^j} = \Gamma_{ij}^{\;\;\;h}\,\de
/de{x^h}
\end{equation}
for the connection coefficients.

In order to evaluate the variation $\delta{\cal A}\/$ under
arbitrary deformations of the connection, we recall that, given a
metric tensor $g_{ij}\/$, every metric connection $\Gamma\/$ may
be expressed as
\begin{equation}\label{00.2}
\Gamma_{ij}^{\;\;\;h} =\tilde{\Gamma}_{ij}^{\;\;\;h} -
K_{ij}^{\;\;\;h}
\end{equation}
where (in the holonomic basis ${\displaystyle \left\{
\frac{\partial}{\partial x^i}, dx^i \right \}}$)
$\tilde{\Gamma}_{ij}^{\;\;\;h}$ denote the coefficients  of the
Levi--Civita connection associated with the metric $g_{ij}\/$,
while $K_{ij}^{\;\;\;h}\/$ indicate the components of a tensor
satisfying the antisymmetry property $K_{i}^{\;\;jh} = -
K_{i}^{\;\;hj}\/$. This last condition ensures the metric
compatibility of the connection $\Gamma\/$.

In view of this, we can identify the actual degrees of freedom of
the theory with the (independent) components of the metric $g\/$
and the tensor $K\/$. Moreover, it is easily seen that the
curvature and the contracted curvature tensors associated with
every connection \eqref{00.2} can be expressed respectively as
\begin{subequations}
\begin{equation}\label{00.3}
R^h_{\;\;iqj}= \tilde{R}^h_{\;\;iqj} +
\tilde{\nabla}_jK_{qi}^{\;\;\;h} -
\tilde{\nabla}_qK_{ji}^{\;\;\;h} +
K_{ji}^{\;\;\;p}K_{qp}^{\;\;\;h} -
K_{qi}^{\;\;\;p}K_{jp}^{\;\;\;h}
\end{equation}
and
\begin{equation}\label{00.4}
R_{ij}=R^h_{\;\;ihj}=\tilde{R}_{ij} +
\tilde{\nabla}_jK_{hi}^{\;\;\;h} -
\tilde{\nabla}_hK_{ji}^{\;\;\;h} +
K_{ji}^{\;\;\;p}K_{hp}^{\;\;\;h} -
K_{hi}^{\;\;\;p}K_{jp}^{\;\;\;h}
\end{equation}
\end{subequations}
where $\tilde{R}^h_{\;\;iqj}\/$ and
$\tilde{R}_{ij}=\tilde{R}^h_{\;\;ihj}\/$ are respectively the
Riemann and the Ricci tensors of the Levi--Civita connection
$\tilde\Gamma\/$ associated with the given metric $g\/$, and
$\tilde\nabla\/$ indicates the Levi--Civita covariant derivative.

Making use of the identities \eqref{00.4}, the action functional
\eqref{00.1} can be written in the equivalent form
\begin{equation}\label{00.5}
{\cal A}\/(g,\Gamma)=\int\sqrt{|g|}f\/(g^{ij}(\tilde{R}_{ij} +
\tilde{\nabla}_jK_{hi}^{\;\;\;h} -
\tilde{\nabla}_hK_{ji}^{\;\;\;h} +
K_{ji}^{\;\;\;p}K_{hp}^{\;\;\;h} -
K_{hi}^{\;\;\;p}K_{jp}^{\;\;\;h}))\,ds
\end{equation}
more suitable for variations in the connection. Taking the metric
$g$ fixed, we have the identifications
$\delta\Gamma_{ij}^{\;\;\;h}=\delta K_{ij}^{\;\;\;h}\/$ and then
the variation
\begin{equation}\label{00.6}
\begin{split}
\delta{\cal A}= \int \sqrt{|g|}f'\/(R)g^{ij}(
\tilde{\nabla}_j\delta{K}_{hi}^{\;\;\;h} -
\tilde{\nabla}_h\delta{K}_{ji}^{\;\;\;h} +
\delta{K}_{ji}^{\;\;\;p}K_{hp}^{\;\;\;h} +
K_{ji}^{\;\;\;p}\delta{K}_{hp}^{\;\;\;h} \\ -
\delta{K}_{hi}^{\;\;\;p}K_{jp}^{\;\;\;h} -
K_{hi}^{\;\;\;p}\delta{K}_{jp}^{\;\;\;h})ds
\end{split}
\end{equation}
Using the divergence theorem, taking the antisymmetry properties
of $K\/$ into account and renaming finally some indexes, we get
the expression
\begin{equation}\label{00.7}
\begin{split}
\delta{\cal A}= \int \sqrt{|g|}\left[ -\de{f'}/de{x^i}\delta^h_j +
\de{f'}/de{x^j}\delta^h_i + f'K_{pj}^{\;\;\;p}\delta^h_i -
f'K_{pi}^{\;\;\;p}\delta^h_j - f'K_{ij}^{\;\;\;h} \right. \\
\left. + f'K_{ji}^{\;\;\;h}\right]\delta{K}_h^{\;\;ij}\,ds
\end{split}
\end{equation}
The requirement $\delta{\cal A}=0\/$ yields therefore a first set
of field equations given by
\begin{equation}\label{00.8}
K_{pj}^{\;\;\;p}\delta^h_i - K_{pi}^{\;\;\;p}\delta^h_j -
K_{ij}^{\;\;\;h} + K_{ji}^{\;\;\;h} =
\frac{1}{f'}\de{f'}/de{x^p}\/\left(\delta^p_i\delta^h_j -
\delta^p_j\delta^h_i\right)
\end{equation}
Considering that the torsion coefficients of the connection
$\Gamma\/$ are $T_{ij}^{\;\;\;h}:= \Gamma_{ij}^{\;\;\;h} -
\Gamma_{ji}^{\;\;\;h} = -K_{ij}^{\;\;\;h} + K_{ji}^{\;\;\;h}\/$
and thus (due to antisymmetry)
$T_{pi}^{\;\;\;p}=-K_{pi}^{\;\;\;p}\/$, eqs.~\eqref{00.8} can be
rewritten as
\begin{equation}\label{00.9a}
T_{ij}^{\;\;\;h} + T_{jp}^{\;\;\;p}\delta^h_i -
T_{ip}^{\;\;\;p}\delta^h_j =
\frac{1}{f'}\de{f'}/de{x^p}\/\left(\delta^p_i\delta^h_j -
\delta^p_j\delta^h_i\right)
\end{equation}
or, equivalently, as
\begin{equation}\label{00.9b}
T_{ij}^{\;\;\;h} = -
\frac{1}{2f'}\de{f'}/de{x^p}\/\left(\delta^p_i\delta^h_j -
\delta^p_j\delta^h_i\right)
\end{equation}
In order to study the variation $\delta{\cal A}\/$ under arbitrary
deformations of the metric, it is convenient to resort to the
representation \eqref{00.1}. Indeed, from the latter, we have
directly
\begin{equation}\label{00.10}
\delta{\cal A}=\int{\sqrt{|g|}\/\left[ f'\/(R)R_{ij} -
\frac{1}{2}f\/(R)g_{ij} \right]\delta{g^{ij}}\,ds}
\end{equation}
thus getting the second set of field equations
\begin{equation}\label{00.11}
f'\/(R)R_{(ij)} - \frac{1}{2}f\/(R)g_{ij}=0
\end{equation}
Of course, one can obtain the same equations \eqref{00.11}
starting from the representation \eqref{00.5} instead of
\eqref{00.1}. In that case, the calculations are just longer.

As a remark concerning eqs. \eqref{00.11}, it is worth noticing
that any connection satisfying eqs. \eqref{00.2} and \eqref{00.9b}
gives rise to a contracted curvature tensor $R_{ij}\/$
automatically symmetric. Indeed, since the tensor $K\/$ coincides
necessarily with the contorsion tensor, namely
\begin{equation}\label{00.12}
K_{ij}^{\;\;\;h} = \frac{1}{2}\/\left( - T_{ij}^{\;\;\;h} +
T_{j\;\;\;i}^{\;\;h} - T^h_{\;\;ij}\right)
\end{equation}
from eqs. \eqref{00.9b} we have
\begin{equation}\label{00.13}
K_{ij}^{\;\;\;h} = \frac{1}{3}\/\left(T_j\delta^h_i -
T_pg^{ph}g_{ij}\right)
\end{equation}
being
\begin{equation}\label{00.14}
T_i := T_{ih}^{\;\;\;h} = -\frac{3}{2f'}\de{f'}/de{x^i}
\end{equation}
Inserting eq. \eqref{00.13} in eq \eqref{00.4},  the contracted
curvature tensor can be represented as
\begin{equation}\label{00.15}
R_{ij} = \tilde{R}_{ij} + \frac{2}{3}\tilde{\nabla}_{j}T_i +
\frac{1}{3}\tilde{\nabla}_hT^hg_{ij} + \frac{2}{9}T_iT_j -
\frac{2}{9}T_hT^hg_{ij}
\end{equation}
The last expression, together with eqs. \eqref{00.14}, entails the
symmetry of the indexes $i$ and $j$. Therefore, in eq.
\eqref{00.11} we can omit the symmetrization symbol and write
\begin{equation}\label{00.16}
f'\/(R)R_{ij} - \frac{1}{2}f\/(R)g_{ij}=0
\end{equation}
Now, considering the trace of the equation \eqref{00.16}, we get
\begin{equation}\label{00.17}
f'\/(R)R  - 2f\/(R)=0
\end{equation}
The latter is an identity automatically satisfied by all possible
values of $R\/$ only in the special case $f\/(R)=\alpha R^2\/$. In
all other cases, eq.\eqref{00.17} represents a constraint on the
scalar curvature $R\/$.

As a conclusion follows that, if $f\/(R)\not = \alpha R^2\/$, the
scalar curvature $R\/$ has to be constant (at least on connected
domains) and coincides with a given solution value of
\eqref{00.17}. In such a circumstance, eqs.\eqref{00.9b} imply
that the torsion $T^{\;\;\;h}_{ij}\/$ has to be zero and the
theory reduces to a $f\/(R)$-theory without torsion. In
particular, we notice that in the case $f\/(R)=R\/$, eq.
\eqref{00.17} yields $R=0\/$ and therefore eqs. \eqref{00.16} are
equivalent to Einstein's equations in empty space $R_{ij}=0\/$. On
the other hand, if we assume $f\/(R)=\alpha R^2\/$, we can have
non--vanishing torsion. In this case, by replacing eq.
\eqref{00.17} in eqs. \eqref{00.9b} and \eqref{00.16}, we obtain
field equations of the form
\begin{subequations}\label{00.18}
\begin{equation}\label{00.18a}
R_{ij} - \frac{1}{4}Rg_{ij}=0
\end{equation}
\begin{equation}\label{00.18b}
T^{\;\;\;h}_{ij}= -\frac{1}{2R}\de{R}/de{x^i}\delta^h_j +
\frac{1}{2R}\de{R}/de{x^j}\delta^h_i
\end{equation}
\end{subequations}
Finally, making use of eq. \eqref{00.15} and the consequent
relation
\begin{equation}\label{00.19}
R= \tilde{R} + 2\tilde{\nabla}_hT^h - \frac{2}{3}T_hT^h
\end{equation}
in eqs. \eqref{00.18}, we can separately point out the
contribution due to the metric and that due to the torsion. In
fact, directly from eqs. \eqref{00.18a} we have
\begin{equation}\label{00.20}
\tilde{R}_{ij} - \frac{1}{4}\tilde{R}g_{ij} =
-\frac{2}{3}\tilde{\nabla}_jT_i +
\frac{1}{6}\tilde{\nabla}_hT^hg_{ij} -\frac{2}{9}T_iT_j +
\frac{1}{18}T_hT^hg_{ij}
\end{equation}
while from the ``trace'' $T_i := T_{ih}^{\;\;\;h} =
-\frac{3}{2R}\de{R}/de{x^i}\/$ of eqs. \eqref{00.18b}, we derive
\begin{equation}\label{00.21}
\de /de{x^i}\/\left( \tilde{R} + 2\tilde{\nabla}_hT^h -
\frac{2}{3}T_hT^h \right) = -\frac{2}{3}\left( \tilde{R} +
2\tilde{\nabla}_hT^h - \frac{2}{3}T_hT^h \right)\/T_i
\end{equation}
Eqs. \eqref{00.20} and \eqref{00.21} are the  coupled field
equations in vacuum for  metric and  torsion in the $f\/(R)=\alpha
R^2\/$ gravitational theory.

\section{Field equations in presence of matter}
The presence of matter is embodied in the action functional
\eqref{00.1} by adding to the gravitational Lagrangian a suitable
material Lagrangian density ${\cal L}_m\/$, namely
\begin{equation}\label{000.1}
{\cal A}\/(g,\Gamma)=\int{\left(\sqrt{|g|}f\/(R) + {\cal
L}_m\right)\,ds}
\end{equation}
Throughout the paper we shall consider material Lagrangian density
${\cal L}_m\/$ not containing terms depending on torsion degrees
of freedom as in \cite{Sotiriou-Liberati2}. The physical meaning
of this assumption will be discussed later. In this case, the
field equations take the form
\begin{subequations}
\begin{equation}\label{000.2a}
f'\/(R)R_{ij} - \frac{1}{2}f\/(R)g_{ij}=\Sigma_{ij}
\end{equation}
\begin{equation}\label{000.2b}
T_{ij}^{\;\;\;h} = -
\frac{1}{2f'\/(R)}\de{f'\/(R)}/de{x^p}\/\left(\delta^p_i\delta^h_j
- \delta^p_j\delta^h_i\right)
\end{equation}
\end{subequations}
where ${\displaystyle \Sigma_{ij}:= -
\frac{1}{\sqrt{|g|}}\frac{\delta{\cal L}_m}{\delta g^{ij}}\/}$
plays the role of the energy--momentum tensor. From the trace of
eq. \eqref{000.2a}, we obtain a fundamental relation between the
curvature scalar $R\/$ and the trace
$\Sigma:=g^{ij}\Sigma_{ij}\/$, which is
\begin{equation}\label{000.3}
f'\/(R)R -2f\/(R) = \Sigma
\end{equation}
(see also \cite{GRGrev} and references therein). In what follows,
we shall systematically suppose that the relation \eqref{000.3} is
invertible and that $\Sigma \not= \const\/$, thus allowing to
express the curvature scalar $R\/$ as a suitable function of
$\Sigma\/$, namely
\begin{equation}\label{000.4}
R=F\/(\Sigma)
\end{equation}
With this assumption in mind, using eqs. \eqref{000.3} and
\eqref{000.4} we can rewrite equations \eqref{000.2a} and
\eqref{000.2b} in the form
\begin{subequations}
\begin{equation}\label{000.5a}
R_{ij} -\frac{1}{2}Rg_{ij}= \frac{1}{f'\/(F\/(\Sigma))}\left(
\Sigma_{ij} - \frac{1}{4}\Sigma g_{ij} \right) -
\frac{1}{4}F\/(\Sigma)g_{ij}
\end{equation}
\begin{equation}\label{000.5b}
T_{ij}^{\;\;\;h} = -
\frac{1}{2f'\/(F\/(\Sigma))}\de{f'\/(F\/(\Sigma))}/de{x^p}\/\left(\delta^p_i\delta^h_j
- \delta^p_j\delta^h_i\right)
\end{equation}
\end{subequations}
Moreover, making use of eqs. \eqref{00.15} and \eqref{00.19}, in
eq. \eqref{000.5a} we can decompose the contracted curvature
tensor and the curvature scalar in their Christoffel and torsion
dependent terms, thus getting an Einstein--like equation of the
form
\begin{equation}\label{000.6}
\begin{split}
\tilde{R}_{ij} -\frac{1}{2}\tilde{R}g_{ij}= \frac{1}{f'\/(F\/(\Sigma))}\left( \Sigma_{ij} - \frac{1}{4}\Sigma g_{ij} \right) - \frac{1}{4}F\/(\Sigma)g_{ij} - \frac{2}{3}\tilde{\nabla}_{j}T_i \\
+ \frac{2}{3}\tilde{\nabla}_hT^hg_{ij} - \frac{2}{9}T_iT_j -
\frac{1}{9}T_hT^hg_{ij}
\end{split}
\end{equation}
Now, setting
\begin{equation}\label{000.7}
\varphi := f'\/(F\/(\Sigma))
\end{equation}
from the trace of eqs. \eqref{000.5b}, we obtain
\begin{equation}\label{000.8}
T_i := T_{ih}^{\;\;\;h} = - \frac{3}{2\varphi}\de\varphi/de{x^i}
\end{equation}
Therefore, substituting in eqs. \eqref{000.6}, we end up with the
final equations
\begin{equation}\label{000.9}
\begin{split}
\tilde{R}_{ij} -\frac{1}{2}\tilde{R}g_{ij}= \frac{1}{\varphi}\Sigma_{ij} + \frac{1}{\varphi^2}\left( - \frac{3}{2}\de\varphi/de{x^i}\de\varphi/de{x^j} + \varphi\tilde{\nabla}_{j}\de\varphi/de{x^i} + \frac{3}{4}\de\varphi/de{x^h}\de\varphi/de{x^k}g^{hk}g_{ij} \right. \\
\left. - \varphi\tilde{\nabla}^h\de\varphi/de{x^h}g_{ij} -
V\/(\varphi)g_{ij} \right)
\end{split}
\end{equation}
where we defined the effective potential
\begin{equation}\label{000.10}
V\/(\varphi):= \frac{1}{4}\left[ \varphi
F^{-1}\/((f')^{-1}\/(\varphi)) +
\varphi^2\/(f')^{-1}\/(\varphi)\right]
\end{equation}
Eqs. \eqref{000.9} may be difficult to solve, neverthless we can
simplify this task finding solutions for a conformally related
metric. Indeed, performing a conformal transformation of the kind
$\bar{g}_{ij} = \varphi g_{ij}\/$, eqs. \eqref{000.9} may be
rewritten in the easier form (see, for example,
\cite{GRGrev,Olmo,German})
\begin{equation}\label{000.10.1}
\bar{R}_{ij} - \frac{1}{2}\bar{R}\bar{g}_{ij} =
\frac{1}{\varphi}\Sigma_{ij} -
\frac{1}{\varphi^3}V\/(\varphi)\bar{g}_{ij}
\end{equation}
where $\bar{R}_{ij}\/$ and $\bar{R}\/$ are respectively the Ricci
tensor and the Ricci scalar curvature associated with the
conformal metric $\bar{g}_{ij}\/$.

Concerning the connection $\Gamma\/$, solution of the variational
problem $\delta{\cal A}=0\/$, from eqs.~\eqref{00.2},
\eqref{00.13} and \eqref{000.8}, one gets the explicit expression
\begin{equation}\label{000.11}
\Gamma_{ij}^{\;\;\;h} =\tilde{\Gamma}_{ij}^{\;\;\;h} +
\frac{1}{2\varphi}\de\varphi/de{x^j}\delta^h_i -
\frac{1}{2\varphi}\de\varphi/de{x^p}g^{ph}g_{ij}
\end{equation}
We can now compare our results with those obtained for $f(R)\/$
theories in the Palatini formalism
\cite{allemandi,palatinifR,ACCF,Sotiriou,Sotiriou-Liberati1,Sotiriou-Liberati2,Olmo}.
If both the theories (with torsion and Palatini--like) are
considered as ``metric'', in the sense that the dynamical
connection $\Gamma\/$ is not coupled with matter ${\displaystyle
\left(\frac{\delta{\cal L}_m}{\delta\Gamma}=0\right)\/}$ and it
does not define parallel transport and covariant derivative in
space--time, then the two approaches are completely equivalent.
Indeed, in the ``metric'' framework, the true connection of
space--time is the Levi--Civita one associated with the metric
$g\/$ and the role played by the dynamical connection $\Gamma\/$
is just to generate the right Einstein--like equations of the
theory. Now, surprisingly enough, our field equations
\eqref{000.9} are identical to the Einstein--like equations
derived within the Palatini formalism  \cite{Olmo}.

On the other hand, if the theories are genuinely metric--affine,
then they are different even though the condition ${\displaystyle
\frac{\delta{\cal L}_m}{\delta\Gamma}=0\/}$ holds. In order to
stress this point, we recall that in a metric--affine theory the
role of  dynamical connection is not only that of generating
Einstein--like field equations but also defining parallel
transport and covariant derivative in space--time. Therefore,
different connections imply different space--time properties. This
means that the geodesic structure and the causal structures could
not obviously coincide. For a discussion on this point see
\cite{ACCF}. Furthermore, it can be  easily shown that the
dynamical connection \eqref{000.11} differs from that derived
within the Palatini formalism. Indeed the latter results to be the
Levi--Civita connection $\bar{\Gamma}\/$ associated with the
conformal metric $\bar{g}= \varphi g\/$ \cite{Sotiriou,Olmo},
while clearly \eqref{000.11} is not. More precisely,
\eqref{000.11} is related to $\bar{\Gamma}\/$ by the projective
transformation
\begin{equation}\label{000.12}
\bar{\Gamma}_{ij}^{\;\;\;h} =\Gamma_{ij}^{\;\;\;h} +
\frac{1}{2\varphi}\de\varphi/de{x^i}\delta^h_j
\end{equation}
which is not allowed in the present theory because, for a fixed
metric $g$, the connection \eqref{000.12} is no longer metric
compatible.

To conclude, we notice that eqs. \eqref{000.10.1} are deducible
from an Einstein--Hilbert like action functional only under
restrictive conditions. More precisely, let us suppose that the
material Lagrangian depends only on the components of the metric
and not on its derivatives as well as that the trace
$\Sigma=\Sigma_{ij}g^{ij}\/$ is independent of the metric and its
derivatives. Then, from the identities
\begin{equation}\label{000.10.2}
\sqrt{|\bar{g}|}=\varphi^2\sqrt{|g|}, \quad \de
/de{g^{ij}}=\frac{1}{\varphi}\de /de{\bar{g}^{ij}} \quad {\rm and}
\quad  \Sigma_{ij}=-\frac{1}{\sqrt{|g|}}\frac{\delta{\cal
L}_m}{\delta g^{ij}}=-\frac{1}{\sqrt{|g|}}\de{{\cal
L}_m}/de{g^{ij}}
\end{equation}
we have the following relation
\begin{equation}\label{000.10.3}
\Sigma_{ij} = - \varphi\frac{1}{\sqrt{|\bar{g}}|}\de{{\cal
L}_m}/de{\bar{g}^{ij}}:= \varphi\bar{\Sigma}_{ij}
\end{equation}
In view of this, and being $\varphi=\varphi\/(\Sigma)\/$, it is
easily seen that eqs. \eqref{000.10.1} may be derived by varying
with respect to $\bar{g}^{ij}\/$ the action functional
\begin{equation}\label{000.10.4}
{\bar{\cal A}}\/(\bar{g})=\int{\left[\sqrt{|\bar{g}|}\left(\bar{R}
-\frac{2}{\varphi^3}V\/(\varphi) \right)+ {\cal L}_m\right]\,ds}
\end{equation}
Therefore, under the stated assumptions, $f(R)\/$ gravity with
torsion in the metric framework is conformally equivalent to an
Einstein--Hilbert like theory.

\section{Equivalence with scalar-tensor  theories}
The above considerations directly lead to study the relations
between $f(R)\/$ gravity with torsion and scalar-tensor theories
with the aim to  investigate their possible equivalence. To this
end, we recall that the action functional of a (purely metric)
scalar-tensor theory is
\begin{equation}\label{00001}
{\cal A}\/(g,\varphi)=\int{\left[\sqrt{|g|}\left(\varphi\tilde{R}
-\frac{\omega_0}{\varphi}\varphi_i\varphi^i - U\/(\varphi)
\right)+ {\cal L}_m\right]\,ds}
\end{equation}
where $\varphi\/$ is the scalar field which, depending on the sign
of the kinetic term, could assume also the role of a phantom field
\cite{faraphantom},  ${\displaystyle \varphi_i :=
\de\varphi/de{x^i}\/}$ and $U\/(\varphi)\/$ is the potential of
$\varphi\/$. For $U\/(\varphi)=0\/$ such a theory reduces to the
standard Brans--Dicke theory \cite{brans}. The matter Lagrangian
${\cal L}_m\/(g_{ij},\psi)\/$ is a function of the metric and some
matter fields $\psi\/$; $\omega_0\/$ is the so called Brans--Dicke
parameter. The field equations derived by varying with respect to
the metric and the scalar field are
\begin{equation}\label{0000.2}
\tilde{R}_{ij} -\frac{1}{2}\tilde{R}g_{ij}=
\frac{1}{\varphi}\Sigma_{ij} + \frac{\omega_0}{\varphi^2}\left(
\varphi_i\varphi_j  - \frac{1}{2}\varphi_h\varphi^h\/g_{ij}
\right) + \frac{1}{\varphi}\left( \tilde{\nabla}_{j}\varphi_i -
\tilde{\nabla}_h\varphi^h\/g_{ij} \right) -
\frac{U}{2\varphi}g_{ij}
\end{equation}
and
\begin{equation}\label{0000.3}
\frac{2\omega_0}{\varphi}\tilde{\nabla}_h\varphi^h + \tilde{R} -
\frac{\omega_0}{\varphi^2}\varphi_h\varphi^h - U' =0
\end{equation}
where ${\displaystyle \Sigma_{ij}:= -
\frac{1}{\sqrt{|g|}}\frac{\delta{\cal L}_m}{\delta g^{ij}}\/}$ and
${\displaystyle U' :=\frac{dU}{d\varphi}\/}$.

Taking the trace of eq. \eqref{0000.2} and using it to replace
$\tilde R\/$ in eq. \eqref{0000.3}, one obtains the equation
\begin{equation}\label{0000.4}
\left( 2\omega_0 + 3 \right)\/\tilde{\nabla}_h\varphi^h = \Sigma +
\varphi U' -2U
\end{equation}
By a direct comparison, it is immediately seen that for $\omega_0
=-\frac{3}{2}\/$ and ${\displaystyle U\/(\varphi)
=\frac{2}{\varphi}V\/(\varphi)\/}$ (where $V\/(\varphi)\/$ is
defined as in eq. \eqref{000.10}), eqs. \eqref{0000.2} become
formally identical to the Einstein--like equations \eqref{000.9}
for a $f(R)\/$ theory with torsion. Moreover, in such a
circumstance, eq. \eqref{0000.4} reduces to the algebraic equation
\begin{equation}\label{0000.5}
\Sigma + \varphi U' -2U =0
\end{equation}
relating the matter trace $\Sigma\/$ to the scalar field
$\varphi\/$, exactly as it happens for $f(R)\/$ gravity. In
particular, it is a straightforward matter to verify that (under
the condition $f''\not= 0\/$) eq. \eqref{0000.5} expresses {\it
exactly} the inverse relation of \eqref{000.7}, namely
\begin{equation}\label{0000.6}
\Sigma=F^{-1}\/((f')^{-1}\/(\varphi))\qquad \Leftrightarrow \qquad
\varphi=f'\/(F(\Sigma))
\end{equation}
being $F^{-1}\/(X) = f'\/(X)X - 2f\/(X)\/$. In fact we have
\begin{equation}\label{0000.6.2}
U\/(\varphi) =\frac{2}{\varphi}V\/(\varphi)= \frac{1}{2}\left[
F^{-1}\/((f')^{-1}\/(\varphi)) +
\varphi(f')^{-1}\/(\varphi)\right]=
\left[\varphi(f')^{-1}\/(\varphi)
-f\/((f')^{-1}\/(\varphi))\right]
\end{equation}
so that
\begin{equation}\label{0000.6.3}
U'\/(\varphi) = (f')^{-1}\/(\varphi) +
\frac{\varphi}{f''\/((f')^{-1}\/(\varphi))} -
\frac{\varphi}{f''\/((f')^{-1}\/(\varphi))} = (f')^{-1}\/(\varphi)
\end{equation}
and then
\begin{equation}\label{0000.6.4}
\Sigma = - \varphi\/U'\/(\varphi) +2U\/(\varphi) =
f'\/((f')^{-1}\/(\varphi))\/(f')^{-1}\/(\varphi)
-2f\/((f')^{-1}\/(\varphi)) = F^{-1}\/((f')^{-1}\/(\varphi))
\end{equation}
As a conclusion follows that, in the ``metric'' interpretation,
$f(R)\/$ theories with torsion are equivalent to $\omega_0
=-\frac{3}{2}\/$ Brans--Dicke theories.

Of course, the above statement is not true if we regard $f(R)\/$
theories as genuinely metric--affine ones. Nevertheless, also in
this case it is possible to prove the equivalence between $f(R)\/$
theories with torsion and a certain class of Brans--Dicke
theories, namely $\omega_0 =0\/$ Brans--Dicke theories with
torsion \cite{German}.

In this regard, let us consider the action functional
\begin{equation}\label{00007}
{\cal A}\/(g,\Gamma,\varphi)=\int{\left[\sqrt{|g|}\left(\varphi{R}
- U\/(\varphi) \right)+ {\cal L}_m\right]\,ds}
\end{equation}
where the dynamical fields are respectively a metric $g_{ij}\/$, a
metric connection $\Gamma_{ij}^{\;\;\;k}\/$ and a scalar field
$\varphi\/$. As mentioned above, the action \eqref{00007}
describes a Brans--Dicke theory with torsion and parameter
$\omega_0 =0\/$.

The variation with respect to $\varphi\/$ yields the  field
equation
\begin{equation}\label{00008}
R = U'\/(\varphi)
\end{equation}
To evaluate the variations with respect to the metric and the
connection we may repeat exactly the same arguments stated in the
previous discussion for $f(R)$ gravity. Omitting for brevity the
straightforward details, the resulting field equations are
\begin{equation}\label{0000.9}
T_{ij}^{\;\;\;h} = -
\frac{1}{2\varphi}\de{\varphi}/de{x^p}\/\left(\delta^p_i\delta^h_j
- \delta^p_j\delta^h_i\right)
\end{equation}
and
\begin{equation}\label{0000.10}
R_{ij} -\frac{1}{2}Rg_{ij}= \frac{1}{\varphi}\Sigma_{ij} -
\frac{1}{2\varphi}U\/(\varphi)g_{ij}
\end{equation}
Inserting the content of eq. \eqref{00008} in the trace of eq.
\eqref{0000.10}
\begin{equation}\label{0000.11}
\frac{1}{\varphi}\Sigma - \frac{2}{\varphi}U\/(\varphi) + R =0
\end{equation}
we obtain again an algebraic relation between $\Sigma\/$ and
$\varphi\/$ identical to eq. \eqref{0000.5}.

Therefore, choosing as above the potential ${\displaystyle
U\/(\varphi) =\frac{2}{\varphi}V\/(\varphi)\/}$, from
\eqref{0000.5} we get $\varphi=f'\/(F(\Sigma))\/$. In view of
this, decomposing $R_{ij}\/$ and $R\/$ in their Christoffel and
torsion dependent terms, eqs. \eqref{0000.9} and \eqref{0000.10}
become identical to eq. \eqref{000.8} and \eqref{000.9}
respectively. As mentioned previously, this fact shows the
equivalence between $f(R)\/$ theories and $\omega_0 =0\/$--
Brans--Dicke theories with torsion, in the metric--affine
framework. These considerations can be extremely useful in order
to give a geometrical characterization to the Brans--Dicke scalar
field.

\section{Applications to FRW cosmology}
We have seen that the field equations \eqref{000.9} may be recast
in the form \eqref{000.10.1} by performing a conformal
transformation $\bar{g}_{ij}=\varphi g_{ij}\/$. In order to apply
the above considerations to FRW cosmological models, let us
suppose that $\Sigma_{ij}\/$ is the energy--momentum tensor of a
cosmological perfect fluid with a negligible pressure and energy
density $\rho\/$ (dust case), namely
\begin{equation}\label{00000.1}
\Sigma^{ij}=\rho\/U^iU^j
\end{equation}
where $\rho=\rho\/(\tau)\/$ only depends on the cosmic time and
$U^i\/$ is the four velocity of the fluid satisfying the condition
\begin{equation}\label{00000.2}
g_{ij}U^iU^j=-1
\end{equation}
From now on, we shall suppose $\varphi > 0\/$ (a sufficient
condition for this is $f'>0\/$) so that the vector field
${\displaystyle \bar{U}^i:=\frac{U^i}{\sqrt{\varphi}}\/}$
represents the four velocity of the fluid with respect to the
conformal metric $\bar{g}_{ij}\/$, while $\bar{U}_i :=
\bar{U}^j\bar{g}_{ji}=\sqrt{\varphi}U_i\/$ denotes the
corresponding dual relation. In view of this,  the identity
\begin{equation}\label{00000.3}
\frac{1}{\varphi}\Sigma_{ij}= \frac{\rho}{\varphi}\/U_iU_j=
\frac{1}{\varphi^2}\bar{\Sigma}_{ij}
\end{equation}
holds, where we have defined
$\bar{\Sigma}_{ij}=\rho\/\bar{U}_i\bar{U}_j\/$. Consequently eqs.
\eqref{000.10.1} may be rewritten as
\begin{equation}\label{00000.4}
\bar{G}_{ij} = \frac{1}{\varphi^2}\left( \bar{\Sigma}_{ij} -
\frac{1}{\varphi}V\/(\varphi)\bar{g}_{ij} \right)
\end{equation}
where $\bar{G}_{ij}$ is the Einstein tensor in the barred metric.
We look for a FRW solution $\bar{g}_{ij}\/$ of \eqref{00000.4},
being
\begin{equation}\label{00000.5}
d\bar{s}^2 = -dt^2 + a^2\/(t)\left[ d\psi^2 + \chi^2\,d\theta^2 +
\chi^2\sin^2\theta\,d\phi^2 \right]
\end{equation}
 Therefore, once a solution $\bar{g}_{ij}\/$ is found,
also the conformal metric $g_{ij}=\frac{1}{\varphi}\bar{g}_{ij}\/$
(solution of the starting equations \eqref{000.9}) will be of the
FRW form. Indeed, the line element associated to $g_{ij}\/$ is
\begin{equation}\label{00000.6}
ds^2 = -\frac{1}{\varphi\/(t)}\,dt^2 +
\frac{a^2\/(t)}{\varphi\/(t)}\left[ d\psi^2 + \chi^2\,d\theta^2 +
\chi^2\sin^2\theta\,d\phi^2 \right]
\end{equation}
so that, by performing the time variable transformation
\begin{equation}\label{00000.7}
d\tau := \frac{1}{\sqrt{\varphi\/(t)}}\,dt
\end{equation}
it may be expressed as
\begin{equation}
ds^2 = -d\tau^2 + A^2\/(\tau)\left[ d\psi^2 + \chi^2\,d\theta^2 +
\chi^2\sin^2\theta\,d\phi^2 \right]
\end{equation}
with ${\displaystyle A:= \frac{a}{\sqrt{\varphi}}\/}$. The field
equations \eqref{00000.4} reduce to the Friedmann equations
\begin{subequations}
\begin{equation}\label{00000.8a}
3\left(\frac{\dot a}{a}\right)^2 + \frac{3k}{a^2} =
\frac{\rho}{\varphi^2} + \frac{V\/(\varphi)}{\varphi^3}
\end{equation}
and
\begin{equation}\label{00000.8b}
2\frac{\ddot{a}}{a} + \left(\frac{\dot a}{a}\right)^2 +
\frac{k}{a^2} =  \frac{V\/(\varphi)}{\varphi^3}
\end{equation}
\end{subequations}
which are  the cosmological equations arising from our theory.

For the sake of  completeness, let us derive the conservation laws
of the theory. The Bianchi identities of eq. \eqref{00000.4} give
\begin{equation}\label{00000.9}
\bar\nabla_i\/\left( \frac{1}{\varphi^2} \bar{\Sigma}^{ij} -
\frac{1}{\varphi^3}V\/(\varphi)\bar{g}^{ij} \right)=0
\end{equation}
where $\bar\nabla\/$ denotes the covariant derivative with respect
to the Levi--Civita connection associated with $\bar{g}_{ij}\/$.
In FRW metric, eqs. \eqref{00000.9} reduce to the continuity
equation
\begin{equation}\label{00000.10}
\frac{d}{dt}\/\left( \frac{\rho}{\varphi^2}a^3 \right) +
a^3\frac{d}{dt}\/\left( \frac{V\/(\varphi)}{\varphi^3} \right) =0
\end{equation}
which completes the cosmological dynamical system.

\section{Discussion and Conclusions}

$f(R)$ gravity seems a viable approach to solve some shortcomings
coming from GR, in particular problems related to quantization on
curved spacetime and cosmological issues related to early Universe
(inflation) and late time dark components. Besides, the scheme of
GR is fully preserved and $f(R)$ can be considered a
straightforward extension where the gravitational action has not
to be necessarily linear in the Ricci scalar $R$.

In this paper, we have discussed the possibility that also the
torsion field could play an important role in the dynamics being
the ${\cal U}_4$ manifolds the straightforward generalization of
the pseudo--Riemannian manifolds ${\cal V}_4$ (torsionless)
usually adopted in GR.

As discussed above, torsion field, in the metric-affine formalism,
plays  a fundamental role in clarifying the relations between the
Palatini and the metric approaches: it gives further degrees of
freedom which  contribute, together with  curvature degrees of
freedom, to the dynamics. The aim is to achieve a self-consistent
theory where unknown ingredients as dark energy and dark matter
(up to now not detected at a fundamental level) could be
completely "geometrized". Torsion field assumes a relevant role in
presence of standard matter since it allows to establish a
definite equivalence between scalar-tensor theories and $f(R)$
gravity, also in relation to conformal transformations.

Finally, an important point deserves a further discussion in
relation to the above results. Let us consider the cosmological
equation \eqref{00000.8a}. In the {\it lhs}, it is clear that
standard matter $\rho$ and the effective cosmological constant
${\displaystyle \Lambda_{eff}=\frac{V(\varphi)}{\varphi^3}}$ play
two distinct role into the dynamics: their evolution is "tuned" by
the scalar field $\varphi$ (i.e. $f'(R)$). The first term could be
relevant for large scale clustered structures (always involving
baryonic matter and dark matter), the second term can be read as
dark energy. If at present epoch they are ${\displaystyle
\frac{\rho}{\varphi^2}\simeq \frac{V(\varphi^)}{\varphi^3}}$, this
reveals a simple mechanism to explain why we are today observing
$\Omega_{M}\simeq 0.3$ and $\Omega_{\Lambda}\simeq 0.7$.
Furthermore, if the field $\varphi$ at denominator is small (that
is $f'(R)$ is small) this could be the reason why the amounts of
dark energy and dark matter result huge today.

As a toy model, let us take into account the well known
 $f(R)= R+ \alpha{R^2}$ theory where, obviously,
$f'(R) = 1+ 2\alpha{R}$. As above, the matter stress-energy tensor
is $\Sigma_{ij} = \rho U_i U_j$ and then Eq.(\ref{000.3}) becomes
\begin{equation} (1+ 2\alpha R)R -2R - 2\alpha R^2 = -\rho \qquad
\longleftrightarrow \qquad R=\rho\,.\end{equation} We have
\begin{equation} \varphi(\rho) = f'(R(\rho))=
1+2\alpha\rho\end{equation} and then the term ${\displaystyle
\frac{\rho}{\varphi^2}}$ becomes \begin{equation}
\frac{\rho}{(1+2\alpha\rho)^2}\,.
\end{equation}
Let us consider now the potential term
\begin{equation} V(\varphi)=
\frac{1}{4}\left[ \varphi F^{-1}((f')^{-1}(\varphi)) +
\varphi^2(f')^{-1}(\varphi)\right]\,.
\end{equation}
Being $(f')^{-1}(\varphi)=\rho$, one has
\begin{equation}
\frac{1}{4}\varphi^2(f')^{-1}(\varphi)=\frac{1}{4}(1+2\alpha\rho)^2
\frac{1}{2\alpha}(1+2\alpha\rho
-1)=\frac{1}{4}(1+2\alpha\rho)^2\rho\,,
\end{equation}
and considering the relation $F^{-1}(Y)=f'(Y)K -2f(Y)$, it is
\begin{equation}
\frac{1}{4}F^{-1}((f')^{-1}(\varphi))=\frac{1}{4}F^{-1}(\rho)=-\rho\,.
\end{equation}
We have also
\begin{equation}
\frac{1}{4}\varphi F^{-1}((f')^{-1}(\varphi))=
-\frac{(1+2\alpha\rho)\rho}{4}\,,
\end{equation}
and then we conclude that
\begin{equation}
V(\varphi(\rho))= \frac{\alpha\rho^2(1+2\alpha\rho)}{2}
\end{equation}
and
\begin{equation}
\frac{V(\varphi(\rho))}{\varphi^3}=
\frac{\alpha\rho^2}{2(1+2\alpha\rho)^2}\,.
\end{equation}
These  arguments show that the condition ${\displaystyle
\frac{\rho}{\varphi^2}\simeq \frac{V(\varphi^)}{\varphi^3}}$ can
be simply achieved leading to comparable values of $\Omega_M$ and
$\Omega_{\Lambda}$. A detailed discussion of these topic, also in
relation with data,  will be the argument of a forthcoming paper.

\end{document}